\documentclass[twocolumn,prx,showpacs,preprintnumbers,amsmath,amssymb]{revtex4-1}
\usepackage{lmodern}
\usepackage[english]{babel}
\usepackage[utf8]{inputenc}
\usepackage{graphicx}
\usepackage{graphics}
\graphicspath{{figures/}} 
\usepackage{dcolumn}
\usepackage{bm}
\usepackage{mathbbol}
\usepackage{amsmath}
\usepackage{amssymb}
\usepackage{subfigure}
\usepackage{color,soul}
\newcolumntype{L}{>{$}c<{$}} 
\newcommand{\OO}[1]{\overleftrightarrow{#1}}

\begin{document}
\title{Magnetic particles and strings in iron langasite}

\author{Evgenii Barts}
\email{e.barts@rug.nl}
\author{Maxim Mostovoy}
\affiliation{Zernike Institute for Advanced Materials, University of Groningen, Nijenborgh 4, 9747 AG Groningen, The Netherlands}
\date{\today}

\begin{abstract}

Particle-like topological magnetic defects that can propagate in all spatial directions open a new dimension for design of magnetic memory and data processing devices.
We show that three-dimensional magnetic skyrmions  can be stabilized in non-collinear antiferromagnets, such as the Fe-langasite, Ba$_3$TaFe$_3$Si$_2$O$_{14}$.
Spins in the crystallographic unit cell of this material form a 120-degree ordering transformed by competing exchange interactions into a short-period spiral, which in turn forms a basis for complex large-scale magnetic superstructures stabilized by Dzyaloshinskii-Moriya interactions and applied magnetic fields.
We derive an effective continuum model describing modulated states of Fe-langasite at the 100 nm scale and explore its magnetic phases and topological defects.
The order parameter space of this model is similar to that of superfluid $^{3}$He-A and the three-dimensional topological defect is closely related to the Shankar monopole and hedgehog soliton in the Skyrme model of baryons.

\end{abstract}
\maketitle

\section{Introduction}
Topology of defects in ordered states of matter is governed by the order parameter describing spontaneous symmetry breaking at a phase transition~\cite{Mermin1979}. 
As the number of variables required to characterize an ordered state increases, so does the diversity and complexity of topological defects. 
A very rich variety of defects is found in superfluid $^3$He with the order parameter describing orbital momentum, spin and phase of the condensate~\cite{Vollhardt2003,VOLOVIK1990}.

Nontrivial topology of compact defects does not necessarily make them stable:  a competition between interactions with different properties under the scaling transformation, $\bm{x} \rightarrow \Lambda \bm{x}$, is required  to prevent the collapse of the defect~\cite{Derrick1964}.
Thus isolated Skyrmion tubes in chiral magnets with the diameter of 10-100 nm are stabilized by  Dzyaloshinskii-Moriya (DM) interactions~\cite{Moriya1960,Dzyaloshinskii1964} favoring non-collinear spins, which compete with the Zeeman and magnetic anisotropy energy favoring uniform states~\cite{Bogdanov1989,Bogdanov1994,Muhlbauer2009,Yu2010}. 
Small size and high stability of Skyrmion tubes in bulk chiral magnets and magnetic multilayers as well as their dynamics driven by applied electric currents make them promising information carriers in magnetic memory and data processing devices~\cite{Nagaosa2013,Fert2013,Back2020}. 
Even smaller skyrmions have been recently observed in centrosymmetric magnets~ \cite{Kurumaji2019,Hirschberger2019,Khanh2020}, where they are stabilized by magnetic frustration and/or long-ranged interactions between spins mediated by conduction electrons~\cite{Okubo2012,Leonov2015,Hayami2017}.

\begin{figure}[t]
\includegraphics[scale=0.9]{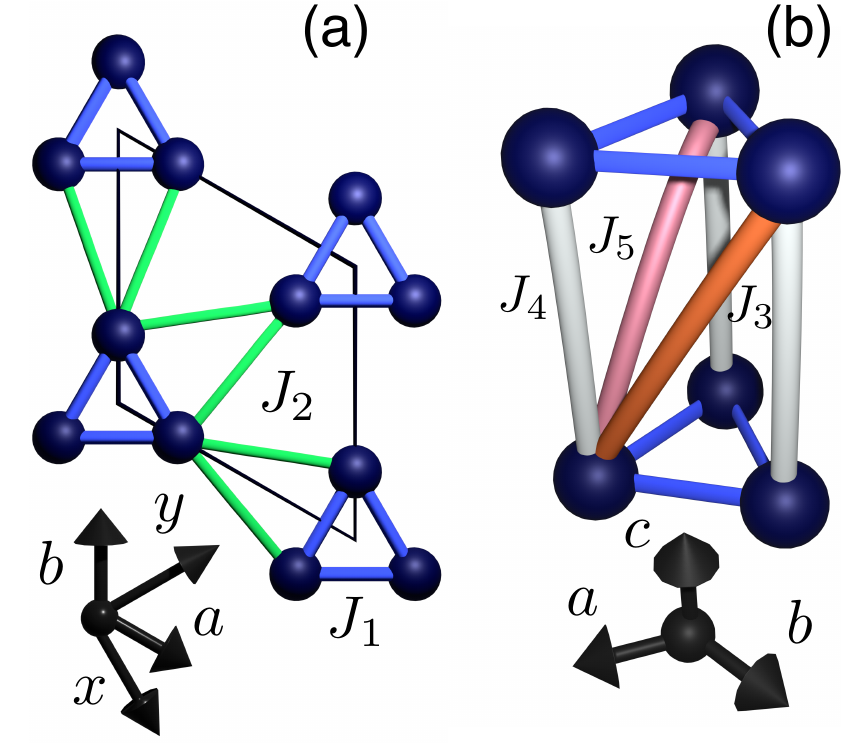}
\caption{(a) Spin triangles in the crystallographic $ab$ plane of  Fe-langasite formed by Fe$^{3+}$ ions (blue spheres). Blue lines mark the bonds between the spins in the triangles with the relatively strong AFM exchange interaction $J_1$. The magnetic trimers form a hexagonal lattice with the AFM coupling $J_2$ between the triangles. (b) Stacking of the spin triangles along the $c$ axis with the competing AFM interactions $J_3, J_4,$ and $J_5$ between the spins in neighboring triangles.}
\label{fig:lattice}
\end{figure}

Here, we discuss a realistic material that can host three-dimensional (3D) magnetic Skyrmions -- non-singular defects that, unlike the Skyrmion tubes, have a finite size in all three spatial directions.
Such particle-like defects can transfer information in all directions, which can stimulate design of novel three-dimensional spintronic devices. 
3D Skyrmions originally emerged as solitons in the non-linear meson model of T.H.R. Skyrme~\cite{Skyrme1962}.
The parameter space of this model, formed by four meson fields, is three-sphere $S^3$ parametrized by three angles. 
A closely related defect, Shankar monopole,  was  predicted to exist (but not yet observed) in the A-phase of superfluid $^3$He~\cite{Volovik1977,Shankar1977}.
The order parameter describing this phase is an SO(3) matrix and the collection of all possible ordered states is projective three-sphere RP$^3$, i.e. S$^3$ with identified diametrically opposite points.
Shankar monopole has been recently realized in the Bose-Einstein condensate of trapped spin-1 particles by application of time-dependent and spatially inhomogeneous magnetic fields~\cite{Lee2018}. 
This defect is, however, unstable and has a short life time.

Higher-dimensional order parameter spaces can also be realized in magnetic materials, in particular, those with triangle-based spin lattices.
Antiferromagnetic (AFM) Heisenberg interactions favor a non-collinear 120$^\circ$ ordering of spins in triangles described by an SO(3) matrix~\cite{Kawamura1984,Dombre1989}.
The three-dimensional order parameter space is formed by the unit vector
$\bm n$ directed along the vector chirality of spins in the triangles~\cite{Kawamura1984} and the angle $\Psi$ describing the spin rotation around ${\bm n}$.
Non-collinear antiferromagnetic orders give rise to electron and magnon bands with non-trivial topology and Weyl fermions~\cite{Chen2014, Kuebler2014, Yang2017, Kuroda2017,Li2020}  resulting in large anomalous Hall and Nernst effects~\cite{Nakatsuji2015,Ikhlas2017} that can be controlled electrically~\cite{Tsai2020}.

We show that 3D skyrmions can naturally occur in the iron langasite, Ba$_3$TaFe$_3$Si$_2$O$_{14}$. 
This fascinating material is both magnetically frustrated and chiral.
The Fe-langasite spin lattice is built of triangles formed by the Fe$^{3+}$-ions ($S = 5/2$) in the $ab$ layers (see Fig.~\ref{fig:lattice}) and the relatively strong AFM exchange interactions result in the $120^\circ$ angle between spins in the triangles~\cite{Marty2008}. 
Furthermore, competing exchange interactions between spins of neighboring triangles, stacked along the $c$ axis, give rise to a helical spiral modulation of the 120$^\circ$-ordering  with the period of $\sim 7$ lattice constants along the $c$ axis. 
The direction of the spin rotation in the spiral and the sign of vector chirality are governed by the chiral nature of the langasite crystal~\cite{Marty2008,Stock2011,Loire2011,Zorko2011} that, otherwise, has little effect on the magnetic ordering. 
However, when the magnetic anisotropy is effectively reduced by an applied  magnetic field, DM interactions give rise to an additional spiral modulation with a long period of  about 2000 \AA\  along a direction parallel to the $ab$ plane~\cite{Ramakrishnan2019}. 
We show that the same DM interactions can stabilize more complex modulated states as well as unusual topological magnetic defects, in particular, particle-like objects carrying  3D Skyrmion topological charge and an associated Hopf number. 

The rest of the paper is organized as follows.
In Sec.~\ref{sec:orderparameter} we discuss the order parameter space of Fe-langasite and in 
Sec.~\ref{sec:model} we derive an effective continuum model describing large-scale variations of the orientation of the spiral plane and spin rotation angle in this material.
The derivation is based on microscopic spin interactions and symmetry analysis. 
The magnetic phase diagram of this model, which includes the experimentally observed field-induced spiral and other  multiply-periodic magnetic superstructures, is discussed in Sec.~\ref{sec:1D}.
Topological defects in two and three spatial dimensions are discussed in sections \ref{sec:2D} and \ref{sec:3D}. 
Technical details are relegated to Appendices.

\section{Order parameter}
\label{sec:orderparameter}
The $120^{\circ}$ order of the classical spins $\bm{S}_1,\bm{S}_2,\bm{S}_3$ in triangles with $\bm{S}_1 + \bm{S}_2 + \bm{S}_3 = 0$ can be described by two orthogonal unit vectors, $\bm{V}_1$ and $\bm{V}_2$ \cite{Reim2018,Ramakrishnan2019}, 
\begin{equation}
\label{eq:SV}
\begin{pmatrix}
\bm{S}_1
\\
\bm{S}_2
\\
\bm{S}_3
\end{pmatrix}
=
\begin{pmatrix}
\bm{V}_1
\\
-\frac{1}{2}\bm{V}_1+\frac{\sqrt{3}}{2}\bm{V}_2
\\
-\frac{1}{2}\bm{V}_1-\frac{\sqrt{3}}{2}\bm{V}_2
\end{pmatrix}.
\end{equation}
The spin 5/2 of Fe$^{3+}$ ions is absorbed into interaction parameters and henceforth $S = 1$. 

Spatial rotations of the frame formed by $\bm{V}_1$, $\bm{V}_2$ and $\bm n = \bm{V}_1 \times \bm{V}_2$ is described by SO(3) matrix $R$ parametrized by three Euler angles, $\phi$, $\theta$ and $\Psi$: 
\begin{equation}
\label{eq:Euler}
\bm{V}_{1,2} = R\bm{V}_{1,2}^{(0)} =  
R_z(\phi)R_y(\theta)R_z(\Psi) \, \bm{V}_{1,2}^{(0)} \, ,
\end{equation}
where $R_z$ and $R_y$ are, respectively, the matrices of rotations around the $z$ and $y$ axes \cite{Edmonds1960}. 
For $\bm{V}_{1}^{(0)}=\bm{e}_x$ and $\bm{V}_{2}^{(0)}=\bm{e}_y$, where $\bm{e}_{x,y}$ are the unit vectors along the corresponding axes, 
\begin{equation}
\begin{split}
\label{eq:V1V2}
&\bm{V}_1=\cos \Psi \, \bm{e}_ \theta + \sin \Psi \, \bm{e}_ \phi   \, ,
\\&
\bm{V}_2=-\sin \Psi \, \bm{e}_ \theta  + \cos \Psi \, \bm{e}_ \phi  \,.
\end{split}
\end{equation}
Here, $\theta$ and $\phi$ are, respectively, the polar and azimuthal angles of the unit vector $\bm{n} = (\sin \theta \cos \phi , \sin \theta \sin \phi , \cos \theta)^T$
and
\begin{equation}
\begin{split}
&
\bm{e}_ \theta 
=\frac{\partial \bm{n}}{\partial \theta}
=(\cos \theta \cos \phi , \cos \theta \sin \phi ,- \sin \theta)^T,\\&
\bm{e}_ \phi
=\frac{1}{\sin \theta}\frac{\partial \bm{n}}{\partial \phi}
=(-\sin \phi, \cos \phi, 0)^T.
\end{split}
\label{eq:ne}
\end{equation}

\begin{figure}[h!]
\includegraphics[scale=1]{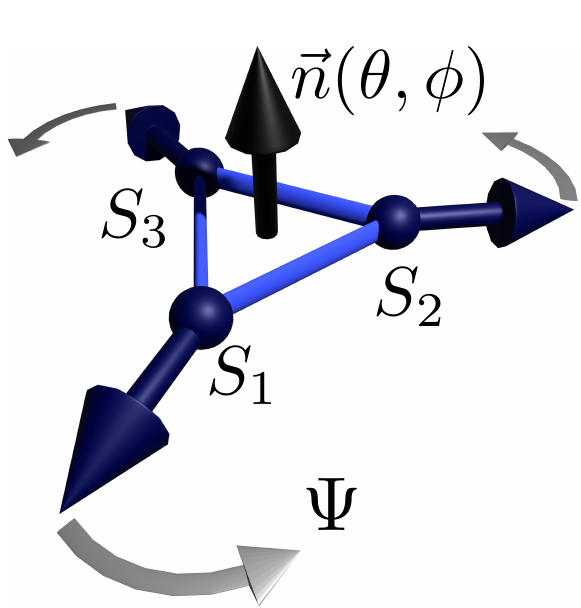}
\caption{The $120^{\circ}$ spin order  parametrized by the polar and azimuthal angles, $\theta$ and $\phi$, describing the  direction of the vector chirality $\bm{n}=
\frac{2}{3\sqrt{3}}\left(
\bm{S}_1 \times \bm{S}_2 
+\bm{S}_2 \times \bm{S}_3 +\bm{S}_3 \times \bm{S}_1 
\right)$, and by the angle $\Psi$,  describing the spin rotation around  $\bm{n}$. }
\label{fig:order}
\end{figure}

The unit vector $\bm{n}(\theta, \phi)=\bm{e}_ \theta \times \bm{e}_ \phi$ describes  the direction of the vector chirality of the $120^{\circ}$ spin order 
(see Fig.~\ref{fig:order}). 

\section{Effective model}
\label{sec:model}

The spiral ordering with the wave vector $Q\|c$, observed in Fe-langasite in zero magnetic field, originates from the competing exchange interactions between the spin triangles stacked along the $c$ direction \cite{Stock2011}:
\begin{equation}
    \tan{Qc} = \sqrt{3} \frac{(J_5-J_3)}{(2J_4-J_3-J_5)}.
\end{equation}
Importantly, the isotropic Heisenberg exchange interactions determine the period of the ‘fast’ spin rotations, $\frac{2 \pi}{|Q|} \sim 7c$, and the sign of $Q$, but not the orientation of the spiral plane described by the vector chirality $\bm n$. 
The latter is governed by much weaker relativistic interactions, namely the $z$-component of DM vectors describing antisymmetric interactions  between spins in the triangles,
\begin{equation}
\label{eq:DMI_tr}
D_z 
\left( \bm{S}_1 \times \bm{S}_2 
+\bm{S}_2 \times \bm{S}_3 +\bm{S}_3 \times \bm{S}_1 \right)_z = \frac{3\sqrt{3}}{2}D_z n_z  ,
\end{equation}
which favors a helical spiral with helicity determined by the sign of $n_z Q$ \cite{Marty2008}, and by an easy-plane magnetic anisotropy favoring in-plane spins and  $n_z = \pm 1$.
The strength of these interactions is two orders of magnitude smaller than the exchange interaction in triangles \cite{Loire2011,Zorko2011,Chaix2016}.
On the other hand, the inter-triangle DM interactions in this chiral magnet tend to induce `slow' variations of $\bm n$ and $\Psi$. Such a magnetic superstructure with the periodicity of $\sim 2000$ \AA\  was observed in Fe-langasite under an  applied magnetic field parallel to the $ab$ plane \cite{Ramakrishnan2019}.
This competition between the magnetic anisotropy favoring a unique direction of $\bm n$ and  the tendency to large-scale modulations, both being relatively weak relativistic effects, can also stabilize topological magnetic defects that are superimposed on the fast spin rotations with the propagation vector along the $c$ direction.

To obtain an effective model describing long-period
magnetic superstructures in Fe-langasite, we separate the fast and slow variations of the order parameter by introducing a slowly varying angle $\psi(\bm r)$: 
\begin{equation}
\label{eq:sep}
\Psi(\bm{r}) = Q z + \psi(\bm{r}),
\end{equation}
The parameter space of the effective model is formed by the three slowly varying angles $\theta$, $\phi$ and $\psi$ (hence, the order parameter  of the effective model is also an SO(3) matrix), which allows us to expand the energy in powers of gradients of these three angles. The energy is then averaged over the fast spin rotations (technical details of the derivation can be found in Appendices A and B).

The energy density of the effective model of the Fe-langasite is,
\begin{equation}
\label{eq:model}
\begin{split}
{\cal E}&=\frac{J_z}{2}\biggl[
(\partial_z \bm{n})^2 +2(D_z \psi)^2
\biggr]
\\&
+\frac{J_\perp}{2}\sum_{\mu=x,y}\biggl[
(\partial_\mu \bm{n})^2 +2(D_\mu \psi)^2
\biggr]\\&
+K_1(1-\cos\theta) + \frac{K_2}{2}(1-\cos^2\theta) - \frac{\chi}{2} (\bm{H} \cdot \bm{n})^2
\\&
+\lambda \biggl[
\cos^2\theta ( -\sin \phi \, \partial_x \theta + \cos\phi \, \partial_y \theta)
+(\bm{n} \cdot \bm{\partial}_\perp)\, \psi
\biggr].
\end{split}
\end{equation}
Here, the first term originates from the interlayer Heisenberg exchange interactions (see Fig.~\ref{fig:lattice}(b)) and  $\displaystyle{J_z = \frac{3}{4} \sqrt{(2J_4-J_3-J_5)^2 +3(J_3-J_5)^2}}$. 
The second term with $\displaystyle{J_\perp = \frac{\sqrt{3}}{2} J_2}$ results from the  exchange interactions between the Fe-triangles in the $ab$ layers (Fig.~\ref{fig:lattice}(a)).
The distances in the direction parallel(perpendicular) to the $c$ axis of the hexagonal lattice are measured in units of the lattice constant, $c$($a$). $D_i \psi = \partial_i \psi + \cos \theta \, \partial_i \phi$ is the covariant derivative of $\psi$ ($i =x,y,z$). This derivative as well as $(\partial_i \bm{n}\cdot\partial_i\bm{n})$ is invariant under an  arbitrary global rotation of spins due to the isotropic nature of Heisenberg exchange interactions. The rotational invariance is equivalent to invariance under $R \rightarrow O R$, where $R$ is defined by Eq.(\ref{eq:Euler}) and $O$ is an arbitrary SO(3) matrix. In terms of the slowly varying variables, invariance of exchange interactions e.g. under  rotation around the $z$ axis through the angle $\alpha$ implies  invariance under $\phi \rightarrow \phi + \alpha$ and invariance under rotation around the $y$ axis through the angle $\beta$ ($|\beta| \ll 1$) implies invariance under $\theta \rightarrow \theta + \beta \cos \phi$, $\phi \rightarrow \phi - \beta \cot \theta \sin \phi$, $\psi \rightarrow \psi + \beta \frac{\sin\phi}{\sin\theta}$.

The third term in Eq.(\ref{eq:model}), playing the role of an internal magnetic field, originates from from DM interactions between spins in the triangles [see Eq.(\ref{eq:DMI_tr})] and the fourth term is the magnetocrystalline anistropy with $K_2 > 0$. Both these terms favor spins parallel to the $ab$ plane and, hence, $\bm n$ parallel or antiparallel to the $c$ axis. The next term is the coupling of the spiral ordering to an applied magnetic field $\bm{H}$, which favors $\bm{n} \parallel\bm{H}$ ($\chi > 0$), since the magnetic susceptibility is the largest for spins rotating in the plane perpendicular to the field vector.  

\begin{table}[htb]
\large
\centering
\begin{tabular}{| L | L | L |}
\hline
 & 3_z & \, 2_y \\
\hline
R_+ & \overline{\omega}  & -\overline{R}_+\\
\overline{R}_+ & \omega  & -R_+\\
R_- & 1  & -\overline{R}_-\\
\overline{R}_- & 1  & -R_-\\
Z & \omega & -\overline{Z}\\
\overline{Z} & \overline{\omega}   & -Z\\
\hline
\end{tabular}
\caption{Symmetry transformation properties of $R_{\pm}$, $Z$ and their complex conjugates (see Eq.(\ref{eq:comp_var})) under the generators of P321 group, $3_z$ and $2_y$. Here, ${\omega=e^{i\frac{2\pi}{3}}}$ and ${\overline{\omega}=e^{-i\frac{2\pi}{3}}}$.} \label{tab:1}
\end{table}

The last term in Eq.(\ref{eq:model}) is a Lifshitz invariant (LI)~\cite{Dzyaloshinskii1964,Bak1980,Bogdanov1989} allowed by the chiral nature of the langasite crystal, $\bm{\partial}_\perp$ being gradient along the in-plane directions. LIs for collinear magnets are expressed in terms of components of the magnetization (or N\'eel) vector and their derivatives. For antiferromagnets with a 120$^\circ$ spin structure described by a more complex order parameter, such as swedenborgaties and langasites,  LIs can be written in terms of the two vectors, $\bm{V}_1$ and $\bm{V}_2$, and their derivatives \cite{Reim2018,Ramakrishnan2019}. 
Finding such LIs is greatly simplified by the use of one-dimensional complex representations of $3_z$. To this end we introduce linear combinations of $\bm{V}_1=(X_1,Y_1,Z_1)$ and 
$\bm{V}_2=(X_2,Y_2,Z_2)$,
\begin{equation}
\label{eq:comp_var}
\begin{split}
&
R_+=X_1+iX_2+i(Y_1+iY_2) = e^{i(\phi-\Psi)} \left(\cos \theta -1\right),
\\&
R_-=X_1+iX_2-i(Y_1+iY_2) = e^{-i(\phi+\Psi)} \left(\cos \theta + 1\right),
\\&
Z=Z_1+iZ_2 = -\sin \theta \, e^{-i\Psi} , 
\end{split}
\end{equation}
and their complex conjugates denoted by $\bar{R}_+$, $\bar{R}_-$ and $\bar{Z}$, respectively. These quantities transform in a simple way under the generators of P321 group, $3_z$ and $2_y$ (Fig.~\ref{fig:lattice}) shown in Table~\ref{tab:1}. These transformation rules follow directly from the symmetry properties of the order parameter:
\begin{equation}
\label{eq:tr_rules}
\begin{split}
3_z :  & \, \phi \to \phi + \frac{2\pi}{3} , \quad \Psi \to \Psi - \frac{2\pi}{3} \, .\\
2_y :  &  \, \phi \to - \phi, \quad \Psi \to \pi -\Psi \, . 
\end{split}
\end{equation}

In this way one obtains 5 LIs favoring an additional modulation with an in-plane wave vector \cite{Ramakrishnan2019}, two of which vanish upon average over the fast degree of freedom. The last term in Eq.(\ref{eq:model}) is the dominant LI resulting from the DM interactions between neighboring triangles in the $ab$ plane (see Appendix~\ref{app:Ani}).
 
\section{Field-induced modulation of the spiral state}
\label{sec:1D}
\begin{figure}[htb]
\includegraphics[scale=0.83]{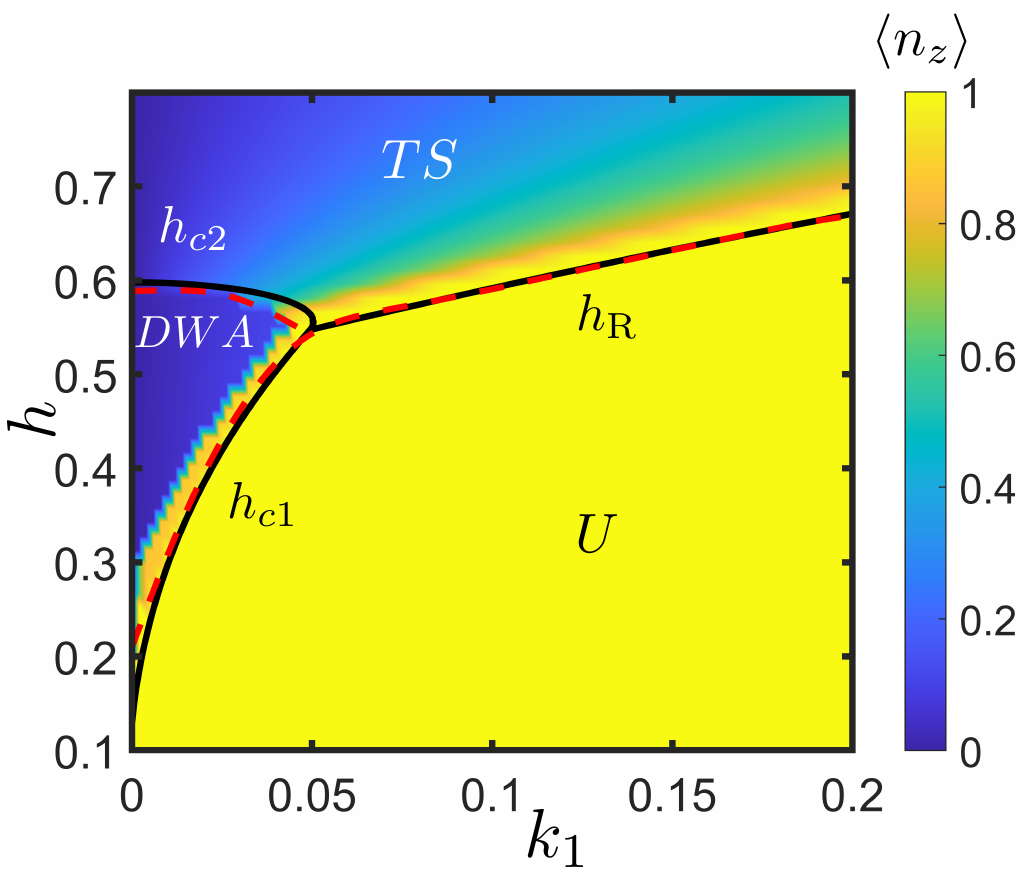}
\caption{Magnetic phase diagram of the model Eq.(\ref{eq:model}) containing the `uniform' (U) spiral state with $\bm{n} \| \hat {\bm z}$ and $\psi = 0$ and the modulated `tilted spiral' (TS) and domain wall array (DWA) states. Solid black(dashed red) phase transition lines are obtained analytically(numerically), the difference being a finite-size effect in numerical calculations. Color indicates average $n_z$. $k_{1,2} = K_{1,2}/ \left(\frac{\lambda^2}{2J}\right)$ and the dimensionless magnetic field, $h$, is defined by $\chi H^2 = h^2  \frac{\lambda^2}{2J}$.  This calculation was performed for $k_2 = 1.25$, $\lambda = 0.4$ and $J=1.0$.}
\label{fig:FD}
\end{figure}
\begin{figure}[htb]
\includegraphics[scale=0.71]{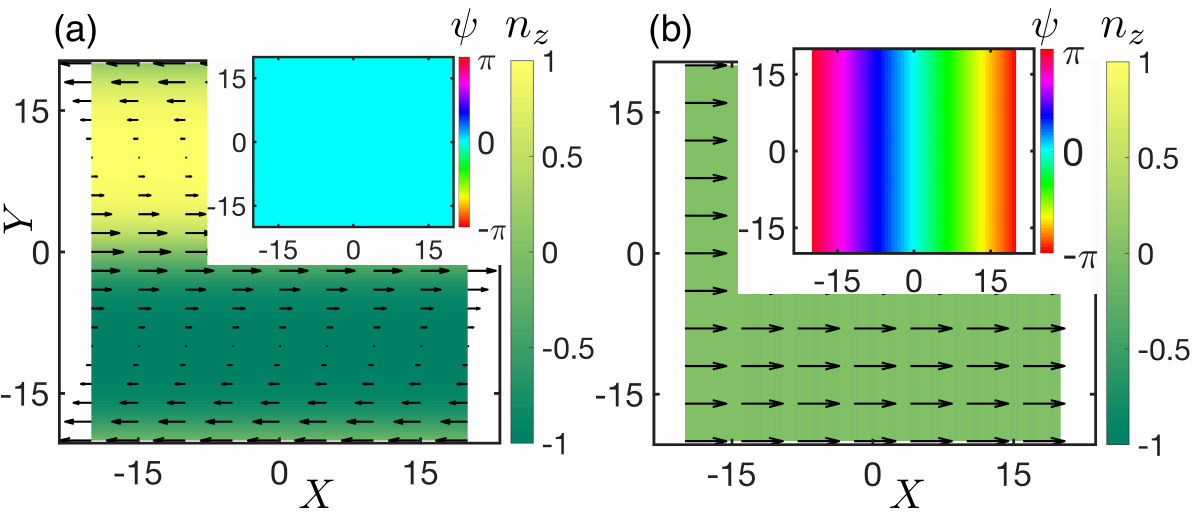}
\caption{Modulated states induced by the magnetic field applied along the $x$ axis: (a) the domain wall array (DWA) state, in which the angle $\theta$ rotates in the $xz$ plane along the $y$ axis normal to the field vector; (b) the `tilted spiral' (TS) state with $\bm{n}\| \bm{H}$ and $\psi$ varying along the field direction. The main figure shows the vector chirality $\bm n$ and the corresponding angle $\psi$ is shown in the inset. In-plane components of $\bm n$ are indicated with arrows; $n_z$ and $\psi$ are color-coded. 
}
\label{fig:gr_states1}
\end{figure}

\begin{figure*}[t]
\includegraphics[scale=1.15]{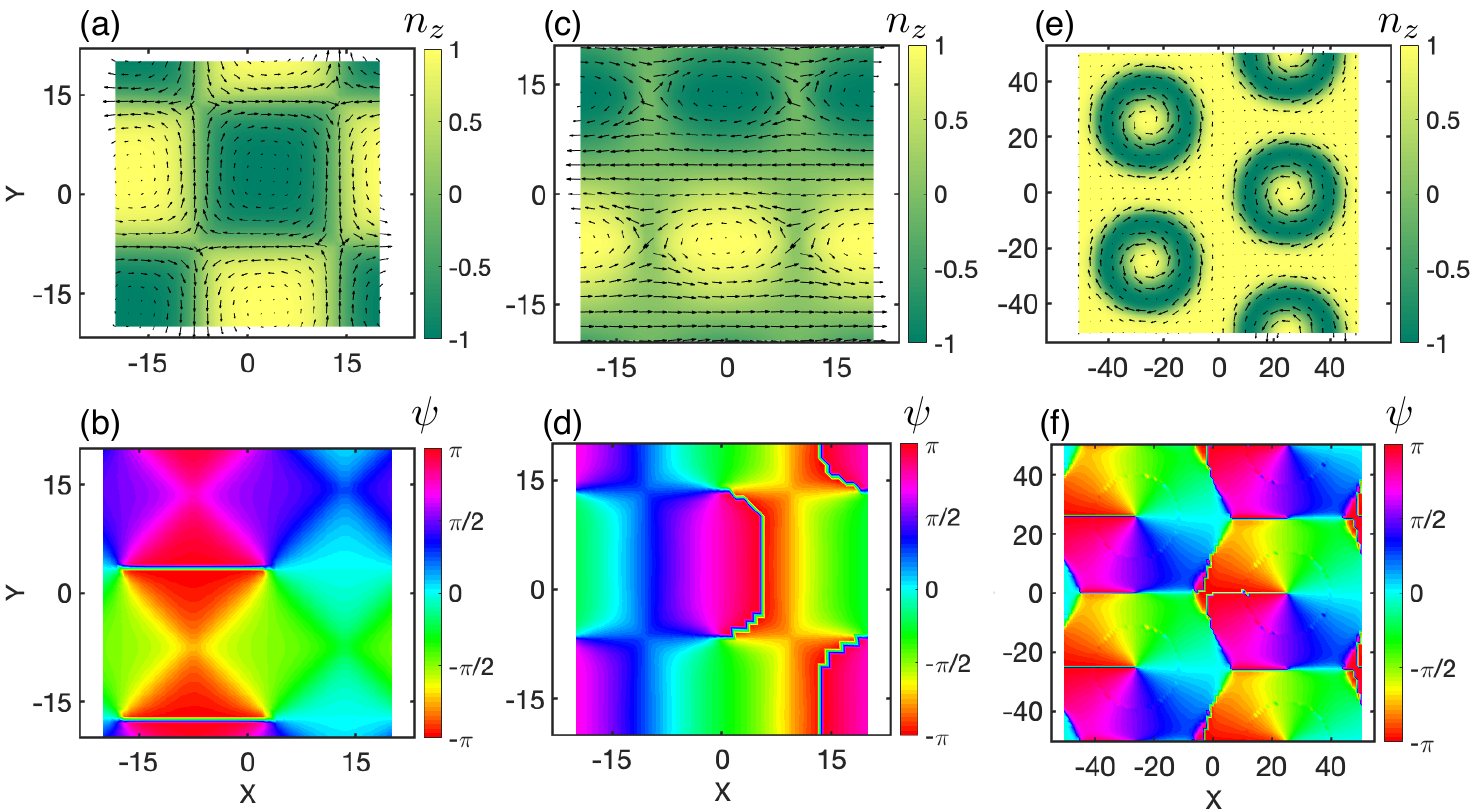}
\caption{Metastable modulated states: (a,b) the vortex array with a square lattice found at low applied magnetic fields ($H<H_{R}$), (c,d) the alternating strings of merons and antimerons metastable at large applied magnetic fields ($H>H_{R}$), and (e,f) the non-singular hexagonal vortex crystal. The first row (panels a,c,e) shows the vector chirality $\bm n$ and the second row (panels b,d,f) shows the corresponding angle $\psi$. In-plane components of $\bm n$ are indicated with arrows; $n_z$ and $\psi$ are color-coded. The angle $\psi$ is plotted modulo $2\pi$  and the lines in the $\psi$-plots are branch cuts, across which $\psi$ discontinuously changes by $2\pi$. }
\label{fig:gr_states2}
\end{figure*}

In zero field, the anisotropy terms with $K_1,K_2 > 0$ confine spins to the $ab$ plane and stabilize the spiral state that can be called `uniform', as in this state $n_z = +1$ and $\psi = {\rm const}$ (in enantiopure samples of Fe-langasite studied in experiments $n_z = -1$ \cite{Marty2008,Stock2011,Loire2011}, which does not affect the phase diagram discussed in this section). An applied magnetic field parallel to the $ab$ plane tends to re-orient the spiral plane, so that it eventually becomes normal to the field  and, hence, $\bm n$ becomes parallel or antiparallel to the field. The re-orientation of $\bm n$, which resembles the spin-flop transition in collinear antiferromagnets, activates LI  and gives rise to additional large-scale modulations. Interestingly, the last term in Eq.(\ref{eq:model}) can stabilize two very different kinds of modulated states: one with a constant $\bm n$ and the angle $\psi$ varying along the field direction, which was recently observed in Fe-langasite above the magnetic field of $\sim 4$ T \cite{Ramakrishnan2019} and another with a constant $\psi$ and $\bm n$ rotating along a direction perpendicular to the field vector. The latter state is similar to the periodic array of domain walls (`mixed state') predicted for collinear chiral antiferromagnets near the flop transition  \cite{Bogdanov1989-2}. Here, we use the effective model Eq.(\ref{eq:model}) to study stable and metastable magnetic states induced by an in-plane magnetic field.

Assuming that in modulated states both $\bm n$ and $\psi$ vary along an in-plane vector $\displaystyle{\bm{\xi} = (\cos \phi_\xi, \sin \phi_\xi,0) }$ (this assumption is verified by numerical simulations), one can exclude $\psi$ from Eq.(\ref{eq:model}) using
\begin{equation}
\label{eq:psi}
\partial_\xi \psi = -\cos \theta \partial_\xi \phi -  \frac{\lambda}{2J_\perp} (\bm{\xi}\cdot \bm{n}).
\end{equation}
The energy density depends then only on $\bm n$:
\begin{equation}
\label{eq:model_sp}
\begin{split}
{\cal E}&=
\frac{J_\perp}{2} (\partial_\xi \bm{n})^2 
+K_1(1-\cos\theta) + \frac{K_2}{2}(1-\cos^2\theta) 
\\&
- \frac{\chi H^2}{2}  n_x^2 -\frac{\lambda^2}{4J_\perp} (\bm{\xi}\cdot \bm{n})^2
-\lambda \sin^2\theta \sin (\phi-\phi_\xi) \, \partial_\xi \theta \, ,
\end{split}
\end{equation}
where $H\|x$. Equation (\ref{eq:model_sp}) resembles the energy of a collinear chiral antiferromagnet~\cite{Bogdanov1989-2}, except for the form of the LI and the `internal magnetic field' $\propto K_1$. 

Figure \ref{fig:FD} shows the phase diagram of the model Eq.(\ref{eq:model_sp}) in the ($K_1,H$) plane. 
In contrast to collinear antiferromagnets, the order parameter $\bm n$ does not abruptly flop, but rotates continuously away from the $z$ axis in the $xz$ plane, for $H > H_{\rm R}$: $\chi H_{\rm R}^2 = |K_1| + K_2 - \frac{\lambda^2}{2J_\perp}$, and the rotation angle is given by 
\begin{equation}
\label{eq:canted}
\cos \theta_{\rm R} = \frac{K_1}{\chi H^2+\frac{\lambda^2}{2J_\perp}-K_ 2}. 
\end{equation}
While $\bm n$ is constant, the angle $\psi$ varies monotonically, $\psi = q (\bm r \cdot \bm \xi)$ with $q = -  \frac{\lambda}{2J_\perp} \sin \theta_{\rm R}$, as follows from Eq.(\ref{eq:psi}) for $\phi = \phi_{\xi} = 0$. This variation  corresponds to an additional rotation of spins around the field vector (see Fig.~\ref{fig:gr_states1} b recently observed in  Fe-langasite~\cite{Ramakrishnan2019}.  The  magnitude of the wave vector $q$ of this state increases as the field strength grows and $\bm n$ approaches the field direction. This `tilted spiral' (TS) state with $\bm n$ tilted away from the $c$ axis and, hence, from the propagation wave vector (almost  parallel to the $c$ axis) has both helical and transverse components.

In another kind of modulated state, the domain wall array (DWA),  $\psi$ is constant, whereas $\bm n$ rotates in the $xz$ plane along the $y$ axis perpendicular to the applied field (see Fig.~\ref{fig:gr_states1} a). For $\lambda > 0$, $\theta$ increases monotonically and $ \phi - \phi_\xi = \frac{\pi}{2}$. This state only appears for relatively small $K_1$ (see Fig.~\ref{fig:FD}). At a critical field, $H_{c1}$, the energy of the domain wall, across which $\theta$ varies by $2\pi$, vanishes, which marks the transition from the uniform $\bm n || \hat{\bm z}$ state to the modulated state. As the magnetic field increases further, the domain wall energy becomes negative and the domain walls form a periodic array with the period that decreases with the field. This state is similar to the `mixed state' in collinear antiferromagnets~\cite{Bogdanov1989-2}, except that in our case $\bm n$ rotates through the angle $2\pi$ across the wall, since the states with $\bm n$ parallel and antiparallel to $\hat{\bm z}$ have different energies for $K_1 \neq 0$. At the second critical field, $H_{c2}$, the transition between the DWA and TS states occurs and  the modulation direction, described by $\bm \xi$, rotates abruptly through $90^\circ$.   

Although the energy of all states in the phase diagram Fig.~\ref{fig:FD} can be found analytically (see Appendix~\ref{app:1Dsim}), we also performed numerical simulations of the model Eq.(\ref{eq:model}) in two spatial dimensions re-written in terms of two orthogonal unit vectors,  $\bm{V}_1$ and $\bm{V}_2$ (see Appendix~\ref{app:2Dsim} for details). Our numerical simulations confirmed the phase diagram Fig.~\ref{fig:FD} and the fact that an additional modulation in all ground states  occurs along one spatial dimension. We also found ``two-dimensional" states -- the vortex array with a square lattice (Fig.~\ref{fig:gr_states2} a,b), the vortex chains  (Fig.~\ref{fig:gr_states2} c,d) and the hexagonal crystal of coreless vortices (Fig.~\ref{fig:gr_states2} e,f). These states are, however, metastable. They may be stabilized by thermal fluctuations at elevated temperatures.

\section{Topological magnetic defects in two spatial dimensions}
\label{sec:2D}

\begin{figure}[htb]
\includegraphics[scale=0.71]{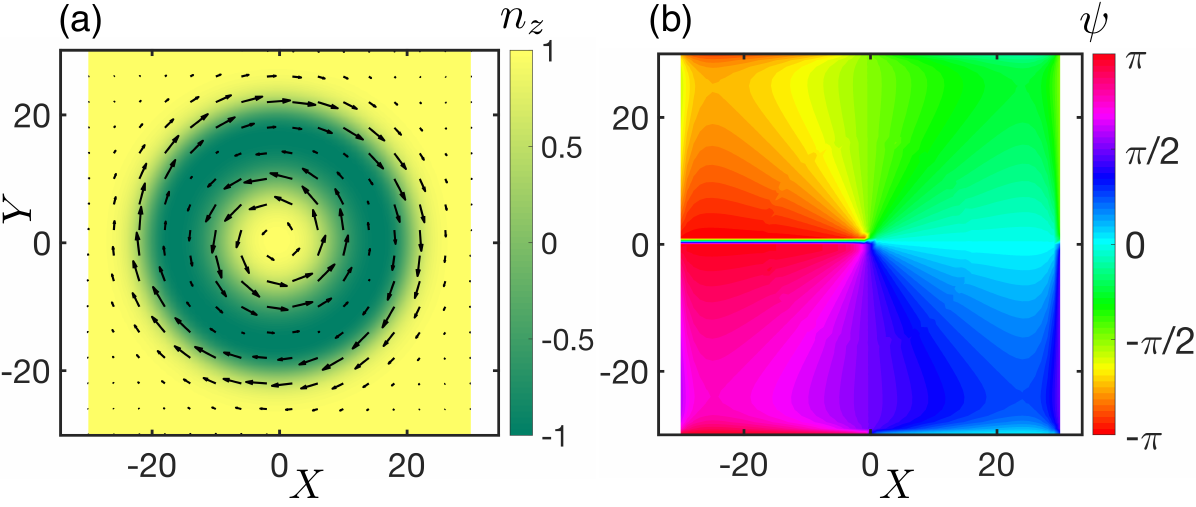}
\caption{
Finite-energy topological defect in the spiral state with $\bm n \| \hat{\bm z}$, which consists of (a) a target skyrmion formed by the unit vector $\bm n$ and (b) a vortex with the angle $\psi$ varying by $- 2 \pi$ along a closed loop around the center of the defect. Arrows show in-plane components of $\bm n$; $n_z$ and $\psi$ are color-coded. The numerical simulation was done for  ${J_\perp=1,} \, {\lambda=0.34,} \, {K_1=0,}\, {K_2=0.1,}$ and $ \bm{H}=0$.
}
\label{fig:topdef}
\end{figure}

Singular topological defects in a model with an SO(3) order parameter in two spatial dimensions -- $Z_2$ vortices with an energy logarithmically diverging with the system size --   have been studied in Ref.~\cite{Kawamura1984}. Here we discuss finite-energy defects which, similarly to magnetic skyrmions, can be classified by topology of $\bm n(x,y)$-textures. However, after the angle $\psi$  is integrated out from Eq.(\ref{eq:model}), the resulting energy functional, $E[\bm n(x,y)]$, becomes non-local: it contains long-ranged Coulomb interactions between the `electric' charges induced by spatial variations of $\bm n$. Below we show that finite-energy textures have zero skyrmion topological charge (magnetic skyrmions are charged and have an infinite `electrostatic' energy) in agreement with $\pi_2({\rm SO(3)}) = 0$.

The electrostatic potential, $\varphi_{\rm el}$, is a variable dual to $\psi$: 
\begin{equation}
D_\mu \psi + \frac{\lambda}{2 J_{\perp}} n_\mu = - \epsilon_{\mu\nu} \partial_\nu \varphi_{\rm el},
\end{equation}
where $\epsilon_{\mu\nu}$ is the antisymmetric tensor ($\mu,\nu = x,y$) and we used the fact that the divergence of the left-hand side is 0, which can be checked by varying the energy Eq.(\ref{eq:model}) with respect to $\psi$. The electrostatic potential satisfies Poisson equation, $- \triangle \varphi_{\rm el} = 4 \pi \rho_{\rm el}$, with the electric charge density
\begin{equation}\label{eq:rhoel}
\rho_{\rm el} = \frac{1}{4\pi}(\bm n \cdot \partial_x \bm n \times \partial_y \bm n) - \frac{\lambda}{8\pi J_\perp} [\bm \nabla \times \bm n]_z,
\end{equation}  
the first term being the skyrmion charge density. Equation (\ref{eq:model}) can then be written in the form,
\begin{equation}
\label{eq:model2D}
{\cal E}=
\frac{J_\perp}{2}\sum_{\mu=x,y}
(\partial_\mu \bm{n})^2 - \lambda n_z [\bm \nabla \times \bm n]_z + U(\bm n) + \frac{1}{2\epsilon} \varphi_{\rm el} \rho_{\rm el},
\end{equation}
where 
$U(\bm n) = K_1(1-\cos\theta) + \frac{\left(K_2-\frac{\lambda^2}{2J_\perp}\right) }{2}\sin^2\theta - \frac{\chi}{2} (\bm{H} \cdot \bm{n})^2$ and the last term is the positive electrostatic energy with the `dielectric' constant $\epsilon = \frac{1}{8\pi J_\perp}$. 

Finite-energy defects have zero total electric charge, 
 \begin{equation}\label{eq:Qel}
 Q_{\rm el} = \int\!\!d^2x \rho_{\rm el}  = Q_{\rm sk} - \frac{\lambda}{8\pi J_\perp} \oint d \bm x \cdot  \bm n = 0,
 \end{equation} 
where $Q_{\rm sk}$ is the skyrmion charge (Eq.(\ref{eq:Qel}) is similar to the Mermin-Ho relation for the circulation of the superfluid velocity in $^3$He-A~\cite{Mermin1976}). Since for a finite-energy defect the integral over the infinite-radius circle in Eq.(\ref{eq:Qel})  is 0, so is the skyrmion charge $Q_{\rm sk}$.

A stable finite-energy defect in the spiral state with $\bm n \| \hat{\bm z}$ is shown in Fig.~(\ref{fig:topdef}). In polar coordinates $(\rho,\varphi)$, $\phi = \varphi + \frac{\pi}{2}$, $\psi = - \varphi$ and $\theta =  \theta(\rho)$ monotonically increases from 0 at $\rho = 0$ to $2\pi$ at $\rho = \infty$. Thus the $\bm n$-configuration [Fig.~(\ref{fig:topdef})a]  is that of a target skyrmion~\cite{Du2013, Leonov2014, Zheng2017} with zero total skyrmion charge and the angle $\psi$ forms a vortex with the winding number $-1$ [Fig.~(\ref{fig:topdef})b]. 

As in the vortices in type-II superconductors, the covariant derivative $D_\mu \psi$ vanishes  far away from the vortex. In contrast to superconductors, it also vanishes at $\rho = 0$, so that the $\psi$-vortex has no core  and a finite energy. Note that $\phi + \psi = {\rm const}$ in the vortex center, where $\theta = 0$, corresponds to non-rotating spins. 

The stabilization of this defect by 
the LI in  Eq.\eqref{eq:model} follows from the fact that this term favors $\psi$ varying along $\bm n$ and $\theta$ varying along the direction normal to $\bm n$. Both these trends are fulfilled in this coreless vortex, in which the in-plane component of $\bm n$ is along the azimuthal direction and $\theta$ varies along the radial direction. Topological protection is ensured by the existence of a non-contractible loop in the SO(3) manifold: $\pi_1({\rm SO(3)}) = Z_2$. A path from the center of the defect to infinity along any radial direction is such a loop. In the center of the defect and at spatial infinity $n_z = +1$, whereas inside the green ring in Fig.~(\ref{fig:topdef})a $n_z$ is negative, corresponding to the local reversal of both the vector chirality of spins in triangles and the spiral helicity determined by the sign of $n_z Q$. The rotational symmetry of the defect turns the calculation of $\theta(\rho)$ into a one-dimensional problem (see Appendix~\ref{app:1Dansatz}).

\section{3D Skyrmion}
\label{sec:3D}
\begin{figure*}[ht]
\includegraphics[scale=1.]{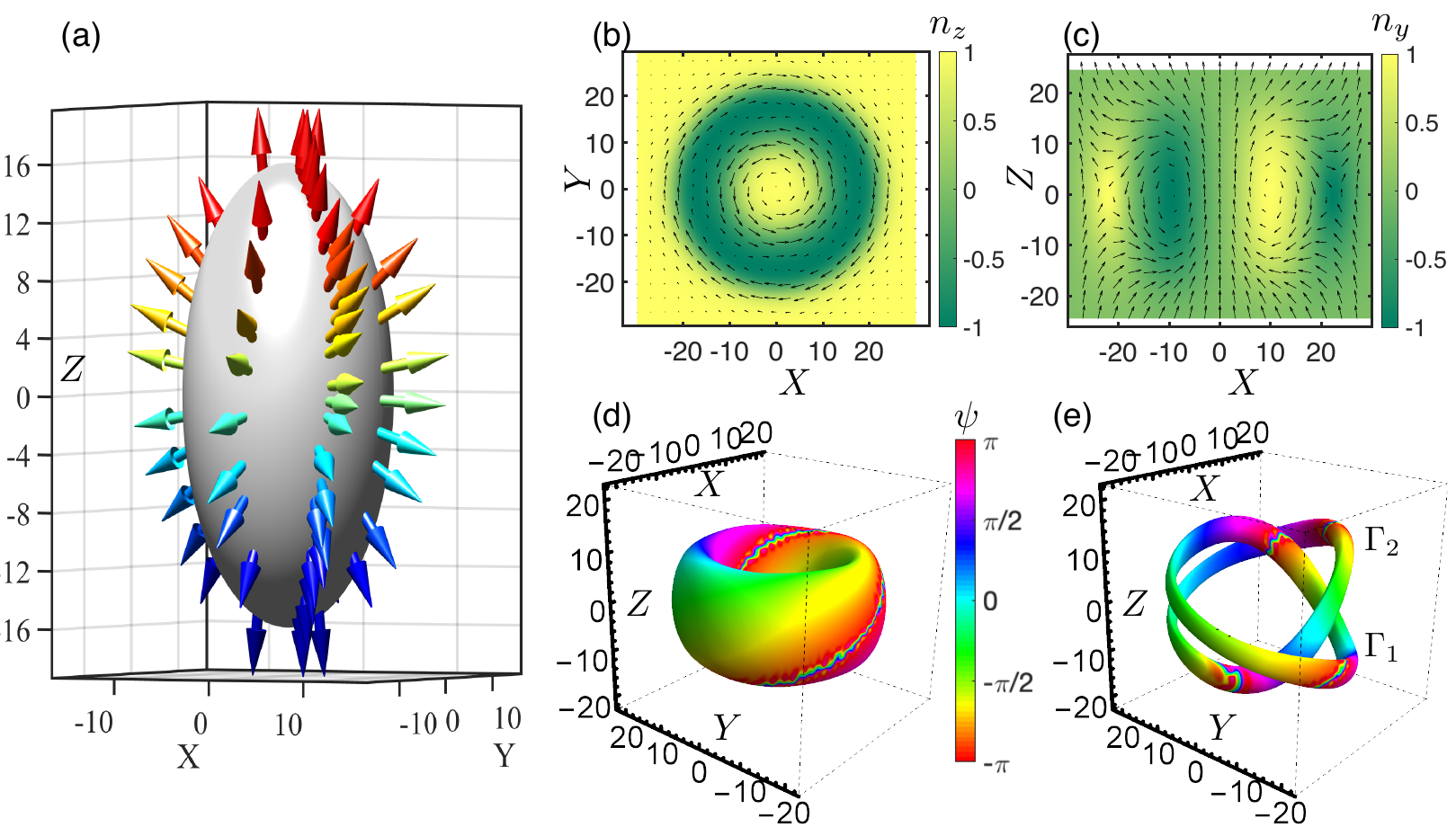}
\caption{
Three-dimensional Skyrmion in a thin layer of Fe-langasite. (a) Arrows indicate the direction of $\bm \Phi =  (\Phi_x,\Phi_y,\Phi_z)$ at the $\Phi_0=0$ surface (grey ellipsoid). The corresponding $\bm{n}$-configuration in the $XY$ plane passing through $Z=0$ (panel b) and in the $XZ$ plane passing through $Y=0$ (panel c). (d) False color plot of the angle $\psi$ at the $n_z = -1/2$ surface. (e) Linking of two closed oriented paths, $\Gamma_{1,2}$, formed by the  constant-$\bm{n}$ lines: $\bm{n} = (0,\sqrt{3}/2,-1/2)^T$, for  $\Gamma_1$, and $\bm{n} = (\sqrt{3}/2,0, -1/2)^T$, for $\Gamma_2$. The numerical simulation was done for ${J_\perp=J_z=1,} \, {\lambda=0.5,} \, {K_1=0,}\, {K_2=0.1}$, $ \bm{H}=0$ and an additional surface anisotropy with  $K_2^s(Z) = 0.26 \cdot \text{exp}(-\frac{(Z_0 - |Z|)}{3.0})$, where $Z_0 = 25$ is half-width of the film.}
\label{fig:3DSkyrm}
\end{figure*}

The third homotopy group, $\pi_3 \left({\rm SO(3)}\right)=Z$, allows for particle-like topological defects that have a finite spatial extent in all three directions.
They are closely related to hedgehog solitons in the non-linear meson model of T.H.R.  Skyrme ~\cite{Skyrme1962} carrying an integer topological charge  ~\cite{Treiman1985},
\begin{equation}\label{eq:topcharge1}
\mathcal{H}=-\frac{1}{96 \pi^2}\int  d^3x \, \varepsilon_{\mu \nu\lambda} \,
{\rm tr}\bigl[
L_\mu L_\nu L_\lambda
\bigr],
\end{equation}
where $\varepsilon_{\mu \nu\lambda}$ is the  antisymmetric Levi-Civita tensor ($\mu,\nu,\lambda = x,y$ or $z$) and $L_{\mu}=U^\dagger \partial_\mu U$,  $U$ being an SU(2) matrix. This matrix is related to the four meson fields,  $(\Phi_0,\Phi_x,\Phi_y,\Phi_z)$, on the 3-sphere,  $\Phi_0^2 + \Phi_x^2 + \Phi_y^2 + \Phi_z^2 = 1$, by $U=\Phi_0 \mathbb{1} + i\bm{\Phi}\cdot\bm{\sigma}$, where $\bm{\sigma} = (\sigma_x,\sigma_y,\sigma_z)$ is a vector composed of Pauli matrices. 
In the isotropic Skyrme's soliton, $\Phi_0$ depends on the radius $r$, varying from $-1$ (the south pole of the 3-sphere) at $r = 0$ to $+1$ (the north pole) at infinity, and the vector $\bm \Phi =  (\Phi_x,\Phi_y,\Phi_z)$ is parallel to the radius vector $\bm r$ (a hedgehog), which guarantees that the 3-sphere formed by the meson fields wraps once around the three-dimensional Euclidean space. 
This topological charge was identified by Skyrme with the baryon number~\cite{Skyrme1962}.

The collapse of Skyrme's baryon in three spatial dimensions is prevented by the interplay between the energy terms of second and fourth order in spatial derivatives of the meson fields.
Terms of fourth order in gradients of the magnetization are relatively small, unless the size of the magnetic defect is as small as one lattice constant.
Two-dimensional Skyrmions (Skyrmion tubes) in chiral magnets are stabilized by DM interactions favoring non-collinear spins.
DM interactions can, in principle,  stabilize 
a 3D topological defect in the chiral Fe-langasite.
Langasite symmetry implies that such defect is axially symmetric. 
Its stability under the scaling transformations along the $z$ and radial directions implies that
\begin{equation}
\label{eq:scaling}
\left\{
\begin{split}
&
E_{\rm ex}^{(z)} = E_{\rm ex}^{(xy)} + E_{\rm DMI} + E_{\rm a} ,
\\ \\
& E_{\rm DMI} +2(E_{\rm ex}^{z} + E_{\rm a})=0,
\end{split}
\right.
\end{equation}
where $E_{\rm ex}^{(z)}$ and $E_{\rm ex}^{(xy)}$ are, respectively, the interlayer and intralayer exchange energies (the first and the second terms in Eq.~\eqref{eq:model}), $E_{\rm DMI}$ is the Lifshitz invariant energy (the last term in Eq.~\eqref{eq:model}) and $E_{\rm a}$ is the energy of the remaining anisotropy terms (the magnetic field $H = 0$).
From Eq.(\ref{eq:scaling}) we obtain, $2E_{\rm ex}^{(xy)}+3E_{\rm DMI}+4E_{\rm a} = 0$, where $E_{\rm DMI}<0$ and $E_{\rm ex}^{(xy)},E_{\rm a}>0$ (the anisotropy energy $E_{\rm a}$ is counted from the energy of the uniform state with $n\|z$).
Although the DMI energy only involves in-plane derivatives of the order parameter, it can make the defect stable both in the radial and $z$ directions provided 
\begin{equation}
\frac{2}{3} E_{\rm ex}^{(xy)} < |E_{\rm DMI}| < 2 E_{\rm ex}^{(xy)},
\end{equation}
where the first inequality results from $E_{\rm a} > 0$ and the second inequality ensures that the total energy of the defect is positive.

Our numerical simulations with periodic boundary conditions in all three directions show that the DM interaction  strength required to stabilize the 3D defect exceeds the critical value, above which the `uniform' spiral state with $n\|z$  and constant $\psi$ becomes unstable towards an additional periodic modulation, i.e. it transforms into the TS or DWA state discussed in Sec.~\ref{sec:1D} in zero magnetic field.
Other symmetry allowed LIs (see Appendix~\ref{app:Ani}) also do not lead to stabilization of the 3D defect in the uniform state.

However, we have found stable 3D defects in slabs with open boundary conditions along the $z$ direction and periodic boundary conditions along the $x$ and $y$ directions (see Fig.~\ref{fig:3DSkyrm}). 
They are stabilized by a surface anisotropy favoring $\bm n$ (anti)parallel to the $c$ axis near the surfaces.
This mechanism is similar to the stabilization of Hopfions in films of liquid crystals by boundary conditions~\cite{Ackerman2016}. 

Figure~\ref{fig:3DSkyrm}a shows that the 3D defect is an axially symmetric hedgehog elongated along the $c$ axis, similar to the spherically symmetric hedgehog in the Skyrme's meson model. 
Here, the grey surface is a surface of zero $\Phi_0$ and the arrows show the direction of $\bm \Phi =  (\Phi_x,\Phi_y,\Phi_z)$ at this surface. The coordinates on the 3-sphere are related to the angles $\theta, \phi$ and $\psi$ by
\begin{equation} \label{eq:S3rep}
\begin{pmatrix}
\Phi_x
\\
\Phi_y
\\
\Phi_z
\\
\Phi_0
\end{pmatrix}
=
\begin{pmatrix}
\sin \frac{\theta}{2} \, \sin \frac{\psi - \phi}{2}
\\
\sin \frac{\theta}{2} \, \cos \frac{\psi - \phi}{2}
\\
\cos \frac{\theta}{2} \, \sin \frac{\psi + \phi}{2}
\\
\cos \frac{\theta}{2} \, \cos \frac{\psi + \phi}{2}
\end{pmatrix},
\end{equation}
$\Phi_0 = -1$ in the center of the 3D skyrmion and $\Phi_0 = +1$ at the periphery, so that the topological charge \eqref{eq:topcharge1} is $-1$. Instead of using the mapping \eqref{eq:S3rep} to  3-sphere, one can calculate topological charge directly by substituting $L_{\mu}$ in 
Eq.(\ref{eq:topcharge1}) with $R^{-1} \partial_\mu R$, where $R \in {\rm SO(3)}$ is defined by Eq.(\ref{eq:Euler}) with $\Psi$ replaced by the slowly varying angle $\psi$ (see Appendix~\ref{app:SKcharge}).

Figure~\ref{fig:3DSkyrm}b shows the $\bm n$-configuration in the $xy$ plane passing through the center of the defect. 
It coincides with that of the 2D target skyrmion (Fig.~\ref{fig:topdef}a). 
On the other hand, the $xz$ cut through the defect (see Fig.~\ref{fig:3DSkyrm}c, where arrows show $x$ and $z$ components of $\bm n$ and $n_y$ is color-coded) shows that the $\bm n$ configuration has a doughnut shape.  
In fact, the $\bm n$-part of the 3D skyrmion is a Hopfion, similar to Hopfions in ferromagnets and liquid crystals~\cite{Cooper1999,Tai2018,Liu2018,Sutcliffe2017,2019arXivRybakov}
and 3D topological charge Eq.(\ref{eq:topcharge1}) equals the Hopf number of the $\bm n$-texture written in terms of the vector potential,
\begin{equation}\label{eq:vecpot}
a_\mu = -D_\mu \psi = \bm{V}_1\cdot\partial_\mu \bm{V}_2, 
\end{equation}
and the corresponding magnetic field $\bm b = [\bm \nabla \times \bm a]$ \cite{Whitehead1947,Kosevich1990},
\begin{equation}\label{eq:topcharge2}
\mathcal{H}=-\frac{1}{16 \pi^2}\int  d^3x \, (\bm a \cdot  \bm b).
\end{equation}
The fact that $n\|z$ both in the center and the outskirts of the defect is not in contradiction with the the opposite signs of $\Phi_0$ in the center and at infinity, since the manifold of SO(3) matrices is obtained by identifying antipodal points of the 3-sphere.

Figure~\ref{fig:3DSkyrm}d shows the false-color plot of the angle $\psi$ at the $n_z = -1/2$ surface (a torus). The angle $\psi$ winds around the torus, which reflects the fact that Hopf number is the linking number for constant-$\bm n$ loops~\cite{Whitehead1947}  (see Fig.~\ref{fig:3DSkyrm}e). The change of the angle $\psi$ along the loop, $\Delta \psi = -\oint d\bm x \cdot \bm a = - 4\pi$. 
Importantly, the 3D Skyrmion is not merely a Hopfion, since the vector chirality $\bm n$ is only a part of the order parameter.
An expression for energy in terms of $\bm n$ only, obtained by integrating out $\psi$ from Eq.(\ref{eq:model}), is non-local: it contains long-ranged interactions between the gradients of $\theta$ and $\phi$.

\section*{Conclusions and outlook}

In conclusion, we studied an effective model describing large-scale spin modulations in Fe-langasite, in particular, the experimentally observed tilted spiral phase. We showed that the three-dimensional order parameter space of this non-collinear antiferromagnet allows for complex spin structures and unconventional topological magnetic defects, such as the coreless vortex tube and three-dimensional Skyrmion. The corresponding vector chirality texture is the target Skyrmion with zero Skyrmion charge, for the tube, and Hopfion with the linking number $\pm 1$, for the 3D Skyrmion. The formal equivalence of the parameter spaces of antiferromagnets with a 120$^\circ$ spin ordering and superfluid $^3$He-A calls for study of magnetic analogs of the wealth of topological defects in the superfluid system~\cite{Vollhardt2003,VOLOVIK1990}.  

It would be interesting to study dynamics of these novel topological magnetic defects induced by electric currents through the spin-transfer and spin-orbit torques. In the Fe-langasite, the defects are embedded into the short-period spiral state, which can give rise to a nontrivial interplay between the collective modes of the spiral and defects. 

There are other antiferromagnets, in which non-collinear spin orders result from frustrated exchange interactions and which can host unusual topological defects, such as manganese nitrides with the cubic inverse perovskite crystal structure showing a variety of non-collinear spin structures and a giant negative thermal expansion effect~\cite{Iikubo2008,Kodama2010,Mochizuki2018}, Pb$_2$MnO$_4$ with a non-centrosymmetric tetragonal crystal lattice and a rare 90$^\circ$ spin ordering ~\cite{Kimber2007},  swedenborgites with alternating triangular and Kagome spin lattices, which similar to Fe-langasite are both frustrated and non-centrosymmetric \cite{Reim2018,Kocsis2016}, and the conducting non-collinear antiferromagnets, MnGe$_3$ and Mn$_3$Sn, showing large anomalous Hall and Nernst effects and allowing for electric control of magnetic states \cite{Chen2014,Kuebler2014,Yang2017, Kuroda2017,Li2020,Liu2017,Nakatsuji2015,Ikhlas2017,Tsai2020}. The unusual  topological defects discussed in this paper can be a new avenue of research in antiferromagnetic spintronics.

\section*{Acknowledgments}
We acknowledge Vrije FOM-programma `Skyrmionics'. 
EB acknowledges insightful discussions with J. Muller, M. Azhar, A. Roy and A. Pozzi. 
We would like to thank the Center for Information Technology of the University of Groningen for their support and for providing access to the Peregrine high performance computing cluster.

\appendix

\section{Effective description of exchange interactions} \label{app:EI}
We first consider the exchange interactions between spin triangles stacked along the $z$ direction parallel to the $c$ axis,
\begin{multline}
E_z =
 \sum_{m,\lambda}\biggl[ J_4\left( \bm{S}_{\lambda, m} \cdot \bm{S}_{\lambda, m+1}\right)+\\
+J_3 \left( \bm{S}_{\lambda, m} \cdot \bm{S}_{\lambda+1, m+1}\right)
+J_5 \left( \bm{S}_{\lambda, m} \cdot \bm{S}_{\lambda-1, m+1}\right)
\biggr],
\end{multline}
where $\lambda=(1,2,3)$ denotes the spin in a triangle ($\bm{S}_{\lambda+3, m} = \bm{S}_{\lambda, m}$) and $m$ is the layer number. The exchange constants $J_3, J_4, J_5 > 0$ describe antiferromagnetic exchange interactions between the spin triangles in neighboring layers (see Fig.1 in the main text). 

In terms of the vectors $\bm{V}_{1,m}$ and $\bm{V}_{2,m}$ describing 
the $120^{\circ}$ spin ordering (see Eq.(1)), the energy reads, 
\begin{equation}
\begin{split}
E_z =&
- \frac{3}{4}\sum_{m} \biggl[
A
\left( \bm{V}_{1,m} \cdot \bm{V}_{1,m+1} + \bm{V}_{2,m} \cdot \bm{V}_{2,m+1} \right)
\\&
+B
\left( \bm{V}_{1,m} \cdot \bm{V}_{2,m+1} - \bm{V}_{2,m} \cdot \bm{V}_{1,m+1} \right)
\biggr],
\end{split}
\end{equation}
where $A=J_3 + J_5 - 2J_4$ and $B=\sqrt{3}(J_5-J_3)$. 
Using Eqs.(\ref{eq:V1V2},\ref{eq:ne}), the energy can then be written in the form,
\begin{equation}
\label{eq:Ezfinal}
    \begin{split}
E_z &=
- \frac{3}{4}\sum_{m} 
\\&
\biggl[
\left( A \cos{(\Psi_{m+1}-\Psi_{m})}
-B \sin{(\Psi_{m+1}-\Psi_{m})}
\right)
\\&
\times
\left( \bm{e}_{\theta,m} \cdot \bm{e}_{\theta,m+1}
+ \bm{e}_{\phi,m} \cdot \bm{e}_{\phi,m+1} \right)
\\&
+ \left(A \sin{(\Psi_{m+1}-\Psi_{m})}
-B \cos{(\Psi_{m+1}-\Psi_{m})} \right)
\\&
\times \left( \bm{e}_{\theta,m} \cdot \bm{e}_{\phi,m+1} - \bm{e}_{\phi,m} \cdot \bm{e}_{\theta,m+1} \right)
\biggr].
\end{split}
\end{equation}
Next we get rid of the fast spin rotation by introducing the slowly varying $\psi_m:$ 
$\Psi_{m} = Q c m + \psi_{m}$, where $Q$ defined by $\tan{Q c} =-\frac{B}{A}$ is the wave vector of the fast-rotating spin spiral. 

Finally, we expand Eq.(\ref{eq:Ezfinal}) in powers of gradients of the slowly varying $\psi_m$, $\bm{e}_{\theta,m}$ and $\bm{e}_{\phi,m}$, which gives the first term in the continuum model ~Eq.(\ref{eq:model}).

A similar procedure applied to exchange interactions between nearest-neighbor spin triangles in the $ab$ layers,
\begin{equation}
\begin{split}
E_{xy} = J_2\sum_{\bm{r}} \biggl[&
 \bm{S}_1\left(\bm{r})  \cdot (\bm{S}_2(\bm{r}-\bm{a}-\bm{b})+\bm{S}_3(\bm{r}-\bm{a}-\bm{b})\right)
 \\&
  +\bm{S}_2(\bm{r}) \cdot (\bm{S}_1(\bm{r}+\bm{a})+\bm{S}_3(\bm{r}+\bm{a}))
  \\&
  +\bm{S}_3(\bm{r}) \cdot (\bm{S}_1(\bm{r}+\bm{b})+\bm{S}_2(\bm{r}+\bm{b}))
\biggr],
\end{split}
\end{equation}
where 
$\hat{\bm a} = 
a\left( 
\frac{\sqrt{3}}{2}
\hat{\bm{x}}+
\frac{1}{2}\hat{\bm y}
\right)$ and 
$\hat{\bm b} = a \left(
-\frac{\sqrt{3}}{2}\hat{\bm{x}}+\frac{1}{2}\hat{\bm y}\right)$ are the basis vectors of the hexagonal lattice, $a$ being the in-plane lattice constant,  
gives the second term in ~Eq.(\ref{eq:model}).

\section{Lifshitz invariants}\label{app:Ani}

There are 5 Lifshitz invariants (LIs) favoring in-plane modulations. They can be derived using the procedure outlined in the main text: ${\rm Im}\left(R_+ \OO{\partial_+} R_{-}\right)$, $ {\rm Im}\left(R_+ \OO{\partial_+} \overline{R}_{-}\right)$, ${\rm Im}\left(R_+ \OO{\partial_-} \overline{Z}\right)$,   ${\rm Im}\left(R_- \OO{\partial_+} \overline{Z}\right)$, and ${\rm Im}\left(R_- \OO{\partial_-} Z\right)$,
where $A\OO{\partial_{\pm}}B=A\partial_i B -B\partial_i A$ and $\partial_{\pm} = \partial_x \pm i \partial_y$. In terms of $\bm{V}_{1}$ and $\bm{V}_{2}$, the LIs have the form
\begin{equation}
\begin{split}
&
L_1=X_1 \OO{\partial_x} Y_1-X_2 \OO{\partial_x} Y_2 
-X_1\OO{\partial_y}Y_2 -X_2 \OO{\partial_y}Y_1 ,
\\&
L_2=2Y_2\OO{\partial_x}Z_2 -2X_1 \OO{\partial_y}Z_1 
+(X_2+Y_1)\left( \OO{\partial_x} Z_1 -\OO{\partial_y}Z_2 \right),
\\&
L_3=- X_1\OO{\partial_x}Z_2 - Y_1 \OO{\partial_y}Z_2 
+ X_2\OO{\partial_x}Z_1 + Y_2\OO{\partial_y}Z_1,
\\&
L_4=X_1\OO{\partial_y}Z_1 - Y_1\OO{\partial_x}Z_1 
+ X_2\OO{\partial_y}Z_2 - Y_2\OO{\partial_x}Z_2,
\\&
L_5=
X_1 \OO{\partial_x} X_2-Y_1 \OO{\partial_x} Y_2 -X_1\OO{\partial_y}Y_2 -Y_1 \OO{\partial_y}X_2.
\end{split}
\end{equation}
The last three survive the averaging over the fast spiral rotations along the $c$ axis: 
\begin{equation}
\begin{split}
&
\langle L_3\rangle =\cos \theta\left(
\sin \phi \, \partial_x \theta - \cos \phi \, \partial_y \theta - (\bm n \cdot \bm{\partial}_\perp) \psi
\right),
\\&
\langle L_4\rangle =\cos^2\theta \left(
\sin \phi \, \partial_x\theta - \cos \phi \, \partial_y\theta
\right) - (\bm n \cdot \bm{\partial}_\perp) \psi ,
\\&
\langle L_5\rangle =\sin2\phi \sin\theta \, \partial_x\theta - \cos2\phi \sin^2\theta \, \partial_x \psi 
\\&
\qquad +\cos2\phi \sin\theta \, \partial_y \theta + \sin 2\phi \sin^2\theta \, \partial_y \psi ,
\end{split}
\end{equation}

The strongest interaction of this kind likely originates from the DMI between neighboring triangles in $ab$ layers,
\begin{equation}
\begin{split}
&
\bm{D}_1 \cdot \bm{S}_2(\bm{r}) \times \bm{S}_1(\bm{r}+\bm{a}+\bm{b})
+\bm{D}_2 \cdot \bm{S}_3(\bm{r}) \times \bm{S}_1(\bm{r}+\bm{a}+\bm{b})\\&
+\bm{D}_3 \cdot \bm{S}_1(\bm{r}) \times \bm{S}_3(\bm{r}-\bm{b})
+\bm{D}_4 \cdot \bm{S}_2(\bm{r}) \times \bm{S}_3(\bm{r}-\bm{b})\\&
+\bm{D}_5 \cdot \bm{S}_3(\bm{r}) \times \bm{S}_2(\bm{r}-\bm{a})
+\bm{D}_6 \cdot \bm{S}_1(\bm{r}) \times \bm{S}_2(\bm{r}-\bm{a}),
\end{split}
\end{equation}
We first consider the interaction due to the $y$-component of $\bm D_1$ together with the symmetry-related interactions on the other bonds:
\begin{equation}
\begin{split}&
\bm{D}_1=D_y \hat{\bm{y}}, \, \bm{D}_2=D_y \hat{\bm{y}}, 
\\&
\bm{D}_3=-D_y \hat{\bm{b}}, \, \bm{D}_4=-D_y\hat{\bm{b}}, 
\\&
\bm{D}_5=-D_y \hat{\bm{a}}, \, \bm{D}_6=-D_y \hat{\bm{a}}.
\end{split}
\end{equation}
(the vectors $\bm a, \bm b, \bm x$ and $\bm y$ are defined in Fig.~\ref{fig:lattice}). In the continuum limit, these interactions give $L_4$ included into Eq.(\ref{eq:model}). 
The DMI resulting from $\bm{D}_1=D_x \hat{\bm{x}}$ and the symmetry related terms give 0 in the continuum limit after averaging over the fast rotations, whereas the interactions resulting from $\bm{D}_1=D_z \hat{\bm{z}}$ modify $K_1$.

In addition, there are 3 LIs favoring an additional modulation along the $c$ axis: 
${\rm Im} \left(R_+\OO{\partial_z} \overline{R}_+\right)$, ${\rm Im} \left(R_-\OO{\partial_z} \overline{R}_-\right)$, and ${\rm Im} \left(Z\OO{\partial_z} \overline{Z}\right)$,
 which can also be written in the form
\begin{equation}
\begin{split}
&
L^z_1=X_1 \OO{\partial_z} Y_1 + X_2 \OO{\partial_z} Y_2,
\\&
L^z_2=X_2 \OO{\partial_z} X_1 + Y_2 \OO{\partial_z} Y_1,
\\&
L^z_3=Z_2 \OO{\partial_z} Z_1.
\end{split}
\end{equation}
Averaging over fast rotations, we obtain
\begin{equation}
\label{outZ}
\begin{split}
&
\langle L^z_1 \rangle \simeq (1+\cos^2\theta)\, \partial_z \phi + 2\cos \theta \, \partial_z \psi ,
\\&
\langle L^z_2 \rangle  \simeq 2\cos \theta \, \partial_z \phi + (1+\cos^2\theta) \, \partial_z \psi ,
\\&
\langle L^z_3 \rangle \simeq \sin^2\theta \, \partial_z \psi.
\end{split}
\end{equation}

\section{Field-induced phase transitions} \label{app:1Dsim}
The transition line between the low-field uniform state with $n\|z$ and the high-field DH state in which $\bm n$ rotates continuously in the $xz$ towards the magnetic field $H \| x$,
\begin{equation}
h_{\rm R}^2 = |k_1| + k_2 - 1,
\end{equation}
was obtained in the main text (above Eq.(\ref{eq:canted})). Here $h$ is the dimensionless magnetic field defined by $\chi H^2 = h^2  \frac{\lambda^2}{2J}$ and $k_{1,2}$ are dimensionless anisotropy parameters: $k_{i} = K_{i}  / \left(\frac{\lambda^2}{2J}\right)$, $i = 1,2$.
 
 At the transition line separating the uniform and DWA states the energy of the domain wall, across which $\theta$ increases by $2\pi$ along the $y$ direction ($\phi = \pi/2$), is $0$. An equation for the domain wall is obtained by minimizing Eq.(\ref{eq:model_sp}) and its first integral has the form,
\begin{equation}\label{eq:DW}
\frac{J_\perp}{2} \left(\frac{d \theta}{dy}\right)^2 -  U(\theta)= V ,
\end{equation}
where 
\[
U(\theta) = K_1(1-\cos\theta) + \frac{K_2}{2}(1-\cos^2\theta) 
\]
and $V$ is a constant. At the transition line, $V = 0$ and the domain wall energy is
\begin{equation}
E_{\rm DW} = \int_0^{2\pi} \sqrt{2J_{\perp}U(\theta)}\, d \theta  - \pi \lambda = 0.
\end{equation}
Calculation of the integral gives an equation for the critical curve, $h_{c1}(k_1)$:
\begin{equation}
\sqrt{k_2 - h_{c1}^2} \left(
\sqrt{1+g} + \frac{g}{2} \ln \frac{2+g+2\sqrt{1+g}}{g}
\right)
 = \frac{\pi}{2\sqrt{2}}
\end{equation}
where $g=\frac{|k_1|}{k_2-h_{c1}^2}$.

For $H > H_{c1}$, $V \neq 0$ and the domain wall energy,
\begin{equation}
E_{\rm DW} = \int_0^{2\pi} \sqrt{2J_{\perp}(U(\theta) + V)}\, d \theta  - \pi \lambda - V L < 0,
\end{equation}
where $L$ is domain wall length, which in the DWA state is finite and given by 
\begin{equation}
L = \sqrt{\frac{J_\perp}{2}}
\int_0^{2\pi} \, \frac{d \theta}{\sqrt{U(\theta)+V}}
\end{equation}  
(see Eq.(\ref{eq:DW})). Minimizing the average energy density,  $\frac{E_{\rm DW}}{L}$, with respect to $V$, we obtain
\begin{equation}\label{eq:V1}
\int_0^{2\pi} \sqrt{2J_{\perp}(U(\theta) + V)}\, d \theta  = \pi \lambda 
\end{equation}
and
\begin{equation}
\frac{E_{\rm DW}}{L} = -V.
\end{equation}

At the critical field $H_{c2}$, the average energy density of the DWA state equals that of the DH state:
\begin{equation}\label{eq:V2}
-V = U(\theta_{\rm R}) - \frac{\chi H_{c2}^2}{2}  \sin ^2 \theta_{\rm R},
\end{equation}
where $\theta_{\rm R}$ is given by  Eq.(\ref{eq:canted}). Solving Eq. (\ref{eq:V1}) with $V$ given by Eq.(\ref{eq:V2}), we obtain the $h_{c2}(k_1)$ line. The phase boundaries discussed in this section are plotted with solid black lines in  Fig.~\ref{fig:FD}.

\section{Numerical simulations} \label{app:2Dsim}
For numerical simulations, we re-write the energy of the effective model Eq.(\ref{eq:model}) in terms of the unit vectors, $\bm{V}_1$ and $\bm{V}_2$,
\begin{equation}
\label{eq:modelV1V2}
\begin{split}
\mathcal{E}&=
\frac{J}{2}\biggl[
(\partial_z \bm{V}_1)^2 + (\partial_z \bm{V}_2)^2 
\biggr]
\\&
+
\frac{J_\perp}{2}\sum_{\mu=x,y}\biggl[
(\partial_\mu \bm{V}_1)^2 + (\partial_\mu \bm{V}_2)^2
\biggr]
\\&
+K_1(1-n_z) + \frac{K_2}{2}(1-n_z^2) - \frac{\chi}{2} (\bm{H} \cdot \bm{n})^2
\\&
-\lambda\sum_{i = 1,2}
\biggl[
 V_i^x \partial_y V_i^{z}-
 V_i^y \partial_x V_i^{z}
\biggr]
+J_{\rm ort} \left( \bm{V}_1 \cdot \bm{V}_2 \right)^2
\end{split}
\end{equation}
where $\bm{n} = \bm{V}_1 \times \bm{V}_2$ and 
$\bm{V}_1$ and $\bm{V}_2$ are slowly varying vectors defined by Eq.(\ref{eq:V1V2}) with $\Psi$ replaced by the slowly varying $\psi$, which removes the fast rotation around the $c$ axis.
The term with a large $J_{\rm ort}> 0$ is added to ensure the  orthogonality of $\bm{V}_1$ and $\bm{V}_2$. We then discretize Eq.(\ref{eq:modelV1V2}) and minimize energy by solving two coupled Landau-Lifshitz-Gilbert equations for the unit vectors $\bm{V}_1$ and $\bm{V}_2$ with an artificially large Gilbert damping.

\section{Two-dimensional topological defect}
\label{app:1Dansatz}

In zero applied magnetic field, the 2D topological defect has rotational symmetry: $\theta = \theta(\rho)$,  $\phi = \varphi + \frac{\pi}{2}$ and $\psi = -\varphi$. The energy  of the defect [see Eq.(\ref{eq:model})] counted from the energy of the spiral state with $\bm n \| \hat{\bm z}$ state is
\begin{equation}
\label{eq:energy_polar}
\begin{split}
E &= 2\pi\int \rho \, d\rho \biggr\{ \\&
\frac{J_\perp}{2}\biggl[
\left(\partial_\rho \theta \right)^2 +\frac{1}{\rho^2} \left(\sin^2\theta +2(1-\cos \theta)^2\right)
\biggr]\\&
+K_1(1-\cos\theta) + \frac{K_2}{2}\sin^2\theta 
-\lambda \biggl[
\cos^2 \theta \, \partial_\rho \theta
+\frac{\sin \theta}{\rho} \biggr]
\biggr\}.
\end{split}
\end{equation}

\begin{figure}[h!]
\includegraphics[scale=0.70]{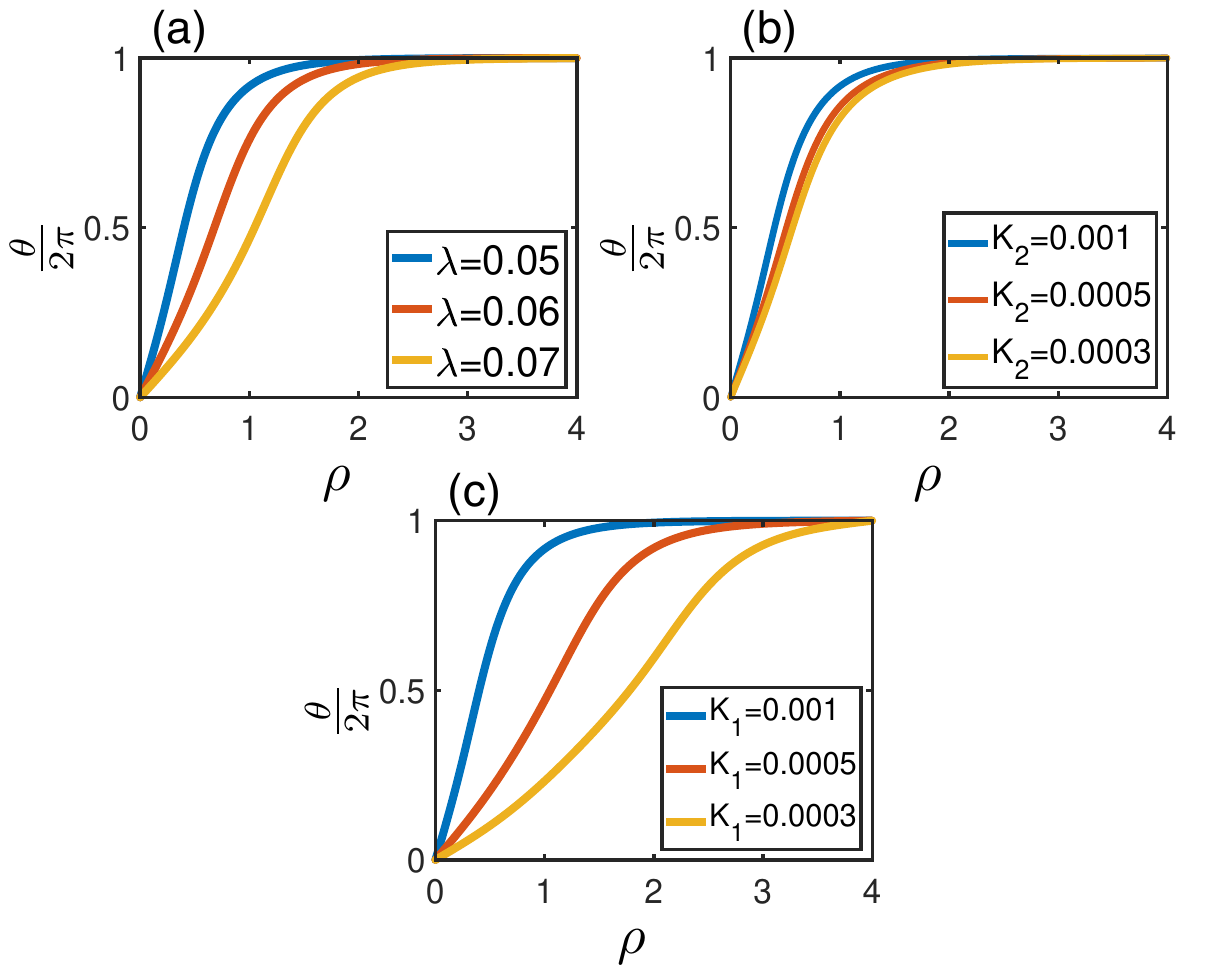}
\caption{$\theta(\rho)$ for (a) three values of $\lambda$ and $K_1=K_2=10^{-3}$, (b) three values of $K_2$, $K_1 = 10^{-3}$ and $\lambda=0.05$, and (c) three values of $K_1$, $K_2 = 10^{-3}$ and  $\lambda=0.05$.
The distance, $\rho$, is measured in units of $\frac{\lambda}{K_1}$.}
\label{fig:topdefth}
\end{figure}

Figure~\ref{fig:topdefth} shows $\theta(\rho)$ found by numerical minimization of Eq.\eqref{eq:energy_polar} with the boundary conditions, $\theta(0) = 0$ and $\theta(\infty) = 2\pi$, for various values of $K_1$ and $K_2$. The soliton radius is independent of $J_{\perp}$ due to the scaling invariance of the exchange energy in two spatial dimensions and  is determined by the relative strength of $\lambda$ and anisotropy parameters.

\section{3D skyrmion topological charge}
\label{app:SKcharge}
Here we show that Eqs.(\ref{eq:topcharge1}) and (\ref{eq:topcharge2}) give the same degree of mapping classified by $\pi_3 ({\rm SO(3)}) = Z$. 
We use
\begin{equation}\label{eq:decomposition}
L_\mu= R^{-1} \partial_\mu R =\sum_{a}A_\mu^a S^a,
\end{equation}
where $S^a$ ($a=x,y,z$) are generators of rotations around the three axes:
\begin{equation}
S^x=
\begin{pmatrix}
0&0&0
\\
0&0&-1
\\
0&1&0
\end{pmatrix}, \,
S^y =
\begin{pmatrix}
0&0&1
\\
0&0&0
\\
-1&0&0
\end{pmatrix}, \,
S^z=
\begin{pmatrix}
0&-1&0
\\
1&0&0
\\
0&0&0
\end{pmatrix},
\end{equation}
satisifying 
\begin{gather}
\bigl[ S^a, S^b\bigr]=\varepsilon^{abc} S^c\, , \\ 
{\rm tr}\left( S^a S^b\right)=-2 \delta^{ab}\, ,\quad   {\rm tr} \left(
S^a S^b S^c
\right)= -\varepsilon^{abc}\, .
\end{gather}

Equation (\ref{eq:topcharge1}) can then be written in the form,
\begin{equation}
\mathcal{H} =\frac{1}{16 \pi^2}\int  d^3x \, \varepsilon_{\mu \nu\lambda} \,
A_\mu^x A_\nu^y A_\lambda^z.
\end{equation}
Substituting the expressions for the expansion coefficients,
\begin{equation}
\begin{split}
&A_\mu^x = \sin \psi \, \partial_\mu \theta - \cos \psi \sin \theta \, \partial_\mu \phi  ,
\\&
A_\mu^y = \cos \psi \, \partial_\mu \theta + \sin \psi \sin \theta \, \partial_\mu \phi ,
\\&
A_\mu^z = \cos \theta \, \partial_\mu \phi - \partial_\mu \psi ,
\end{split}
\end{equation}
we obtain
\begin{equation}
\mathcal{H}=\frac{1}{16\pi^2}\int  d^3x \, 
\varepsilon_{\mu \nu\lambda} \,
\partial_\mu \theta \, \sin \theta \, \partial_\nu \phi \,\partial_\lambda \psi .
\end{equation}

The same expression is obtained by substituting  
$a_i = \bm{V}_1\cdot\partial_i\bm{V}_2 = -D_i \psi =-\partial_i \psi - \cos{\theta} \partial_i \phi$ into Eq.(\ref{eq:topcharge2}).

%
\bibliographystyle{apsrev4-1} %

%

\end{document}